\documentclass[]{spie}  

\usepackage{amsmath,amsfonts,amssymb}
\usepackage{subcaption}
\usepackage{comment}
\usepackage{graphicx}
\usepackage[colorlinks=true, allcolors=blue]{hyperref}

\title{Developing a Wyne Corrector for higher spectral bandwidth focal plane wavefront sensing}

\author[a]{Dominic F. Sanchez}
\author[b]{Benjamin L. Gerard}
\author[b]{Bautista R. Fernandez}
\author[b]{Brian Bauman}
\author[a]{Philip M. Hinz}
\affil[a]{University of California Santa Cruz, Santa Cruz, United States}
\affil[b]{Lawrence Livermore National Laboratory, Livermore, United States}

\begin{document} 
\maketitle

\begin{abstract}
Focal plane wavefront sensing techniques are generally limited to using imaging systems that have below 1\% spectral bandwidths, due to the radial “smearing” of speckles from chromatic diffraction that causes optical image magnification over larger spectral bandwidths. Wyne (1979) designed a pair of triplet lenses to optically minimize this chromatic magnification and increase the spectral bandwidth. Such a Wyne corrector could enable focal plane wavefront sensing at up to 50\% spectral bandwidths and as a result open enable $>50x$ higher-speed focal plane wavefront sensing. We present results of the design and laboratory testing of a Wyne corrector prototype, including a detailed tolerancing analysis considering manufactural wavelength ranges and alignment tolerances. These tests show promising results that this technology can be deployed to current and future high speed focal plane wavefront sensing instruments to enable significant performance enhancements. This document number is LLNL-ABS-857246. 
\end{abstract}


\section{INTRODUCTION}
\label{sec:intro}  
\begin{figure}[b]
    \centering
    \begin{subfigure}[t]{0.65\textwidth}
        \includegraphics[width=1.0\textwidth]{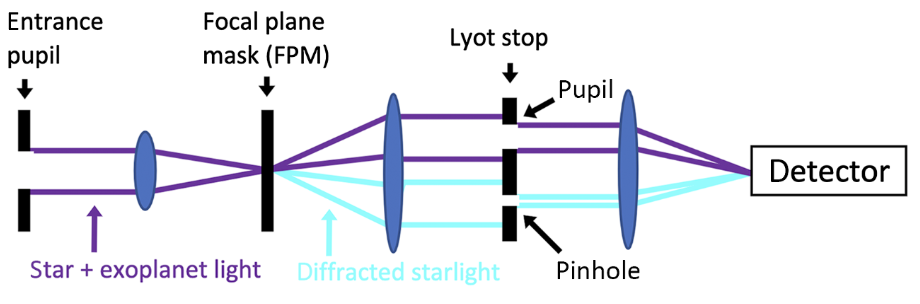}
        \caption{SCC geometric optics-based illustration.}
        \label{fig:scc_intro-a}
    \end{subfigure}
    \begin{subfigure}[t]{0.65\textwidth}
        \includegraphics[width=1.0\textwidth]{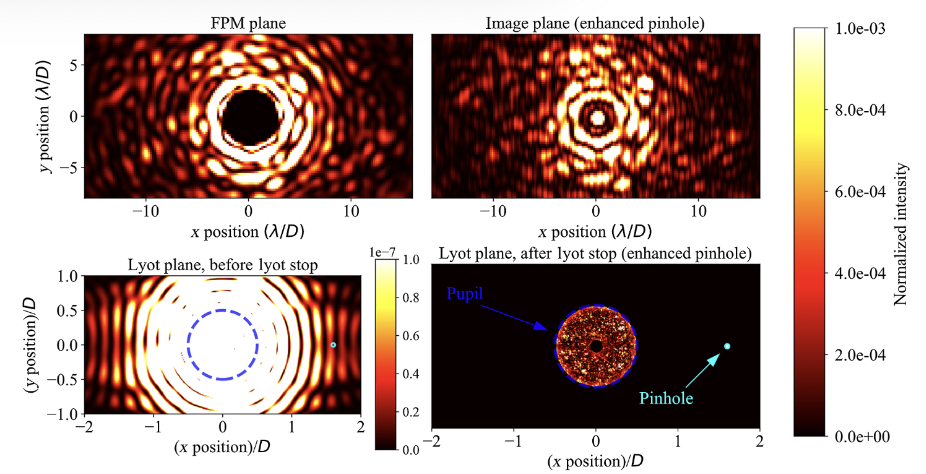}
        \caption{SCC Fourier optics-based illustration.}
        \label{fig:scc_intro-b}
    \end{subfigure}
    \caption{Self-Coherent Camera (SCC) concept descriptions, adapted from Ref. 4}
    \label{fig:scc_intro}
\end{figure}
The Self Coherent Camera (SCC; Fig.~\ref{fig:scc_intro}, Ref.~\citenum{baudoz_self-coherent_2005}.) and other focal plane wavefront sensing techniques generally suffer from decreased performance at higher spectral bandwidths due to radial smearing of speckles. A Wyne corrector (Ref.~\citenum{wyne_extending_1979}) has long been proposed as an optical solution to “unsmear” speckles across larger spectral bandwidths, but never fabricated or tested for this application, until now. The multi-reference SCC (Ref.~\citenum{delorme_focal_2016}) is another solution to enable larger spectral bandwidths, but for $<15\%$ bandwidths, whereas the Wyne corrector can in principle enable $>30\%$. This application is particularly important for high speed focal plane wavefront sensing and control of residual atmospheric speckles where measurement noise dominates the adaptive optics error budget and constrains the accessible loop frames rates for narrow spectral bandwidths (Ref.~\citenum{gerard_laboratory_2022}).

\section{Optical Design}
The fringes produced by an SCC exhibit a small bandwidth due to the coherence condition required for interference. This coherence is wavelength-dependent, and as the wavelength range increases, the fringes begin to smear or blur and causes a reduction in fringe contrast. To increase the effective bandwidth and enhance fringe visibility, the fringes must be scaled to maintain consistent size across the bandwidth.

Given a central reference wavelength $\lambda_{0}$, the reference point spread function (PSF) width can be calculated using:
\begin{equation}
\label{eq:psf_width}
d = 2.44*F{\#}*\lambda_{0} \, ,
\end{equation} 
where $F\#$ is the f-number. The PSF width increases linearly with wavelength; hence, longer wavelengths relative to $\lambda_{0}$ must be magnified and shorter wavelengths must be minified. With a fixed focal length, $F\#$ can be scaled as a function of wavelength by inducing chromatic dispersion and modifying the effective pupil diameter.

Wyne proposed the use of triplet lenses to induce this chromatic dispersion\cite{wyne_extending_1979}. At the central wavelength $\lambda_{0}$, light strikes the first triplet lens at approximately normal incidence, resulting in no optical power contribution. The light then propagates through each cemented element with minimal refraction. This behavior can be understood through the optical power equation,

\begin{equation}
\label{eq:optical_power}
\phi(\lambda) = \frac{n'(\lambda)-n(\lambda)}{R},
\end{equation}
where $R$ is the radius of curvature, and $n$ and $n’$ are the initial and transmitted refractive indices, respectively. When light approaches the first element in the triplet lens, a large $R$ minimizes the optical power contribution. Instead of refracting significantly through the subsequent glass substrates, the light continues to propagate with minimal refraction. The glass substrates for each cemented element in the triplet are chosen to have the same refractive index at the central wavelength $\lambda_{0}$. According to Eq.~\ref{eq:optical_power}, if  $n’(\lambda_{0})=n(\lambda_{0})$, the optical power $\phi(\lambda_{0})$ will be zero. Consequently, these triplet lenses induce little optical power or chromatic magnification at $\lambda_{0}$. However, this condition does not hold for wavelengths shorter or longer than $\lambda_{0}$. The refractive indices of each glass substrate decrease at different rates for these wavelengths, resulting in varying degrees of chromatic dispersion and magnification. At shorter wavelengths relative to $\lambda_{0}$, one glass substrate in the triplet lens has a higher refractive index than the other, while at longer wavelengths, the situation reverses. 

\begin{figure} [ht]
   \begin{center}
   \begin{tabular}{c} 
   \includegraphics[width=12cm]{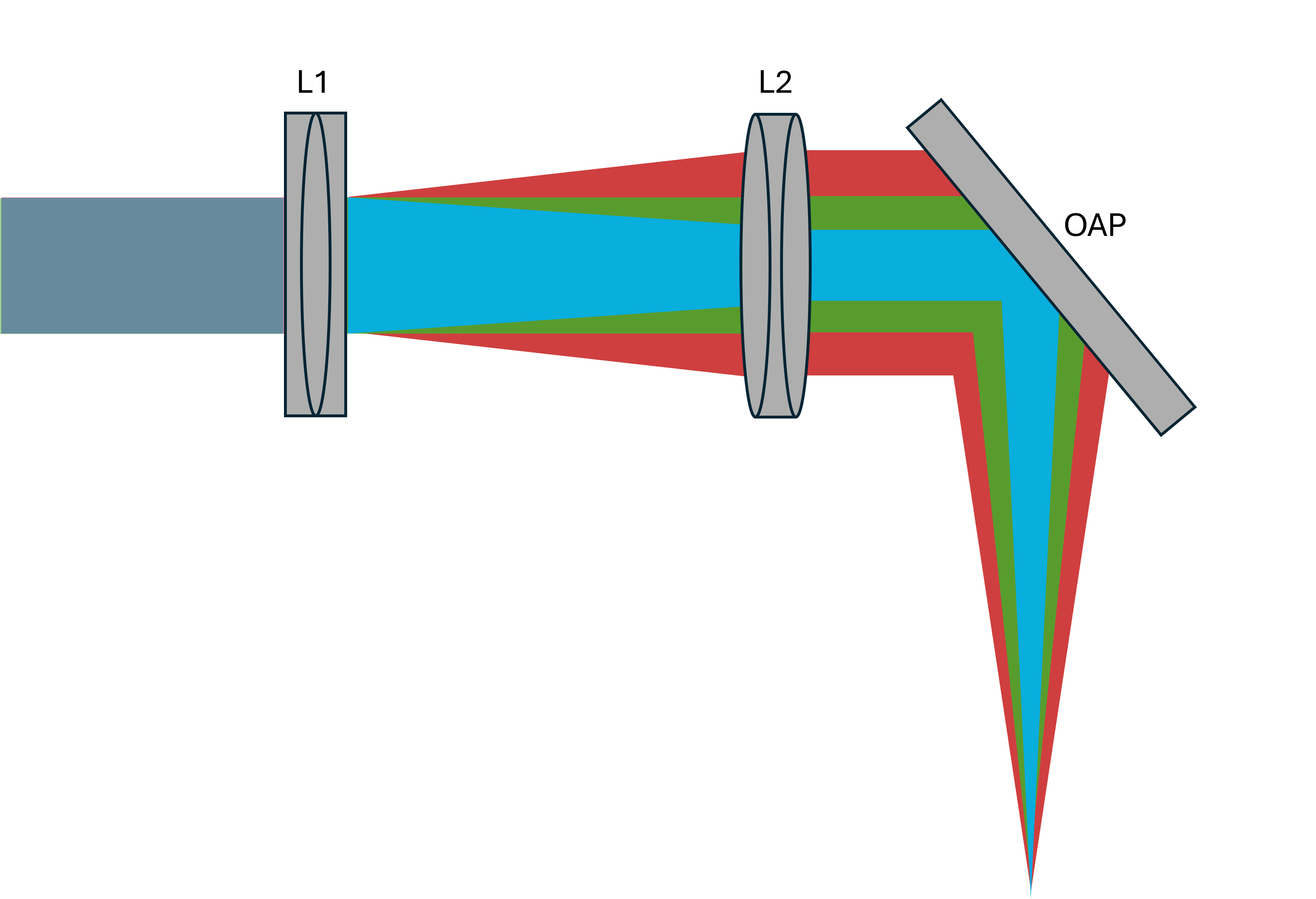}
   \end{tabular}
   \end{center}
   \caption[example] 
   { \label{fig:wynne_concept} Conceptual diagram illustrating the chromatic magnification of an on-axis beam for PSF width matching. The resulting PSF width is ideally constant within the spectral bandwidth and increases fringe visibility.}
   \end{figure}

This change in refractive index induces chromatic dispersion, causing redder light to diverge and bluer light to focus. The second triplet lens then re-collimates this light, chromatically scaling the pupil and adjusting $F\#$ to match the spot size at the central wavelength, as illustrated in Fig.~\ref{fig:wynne_concept}. The amount of chromatic magnification can be controlled by adjusting the separation between the two triplet lenses, as the marginal ray slopes are linear. This linear relationship allows for precise tuning of the chromatic scaling effect. It is important to note that Wyne correctors are particularly advantageous for the SCC because the light source is on-axis. If the source were off-axis, the triplet lenses would induce lateral chromatic aberration, resulting in radial blurring of off-axis sources from the optical axis.

\section{Results}
\label{sec:sections}
\begin{figure} [ht]
   \begin{center}
   \begin{tabular}{c} 
   \includegraphics[width=10cm]{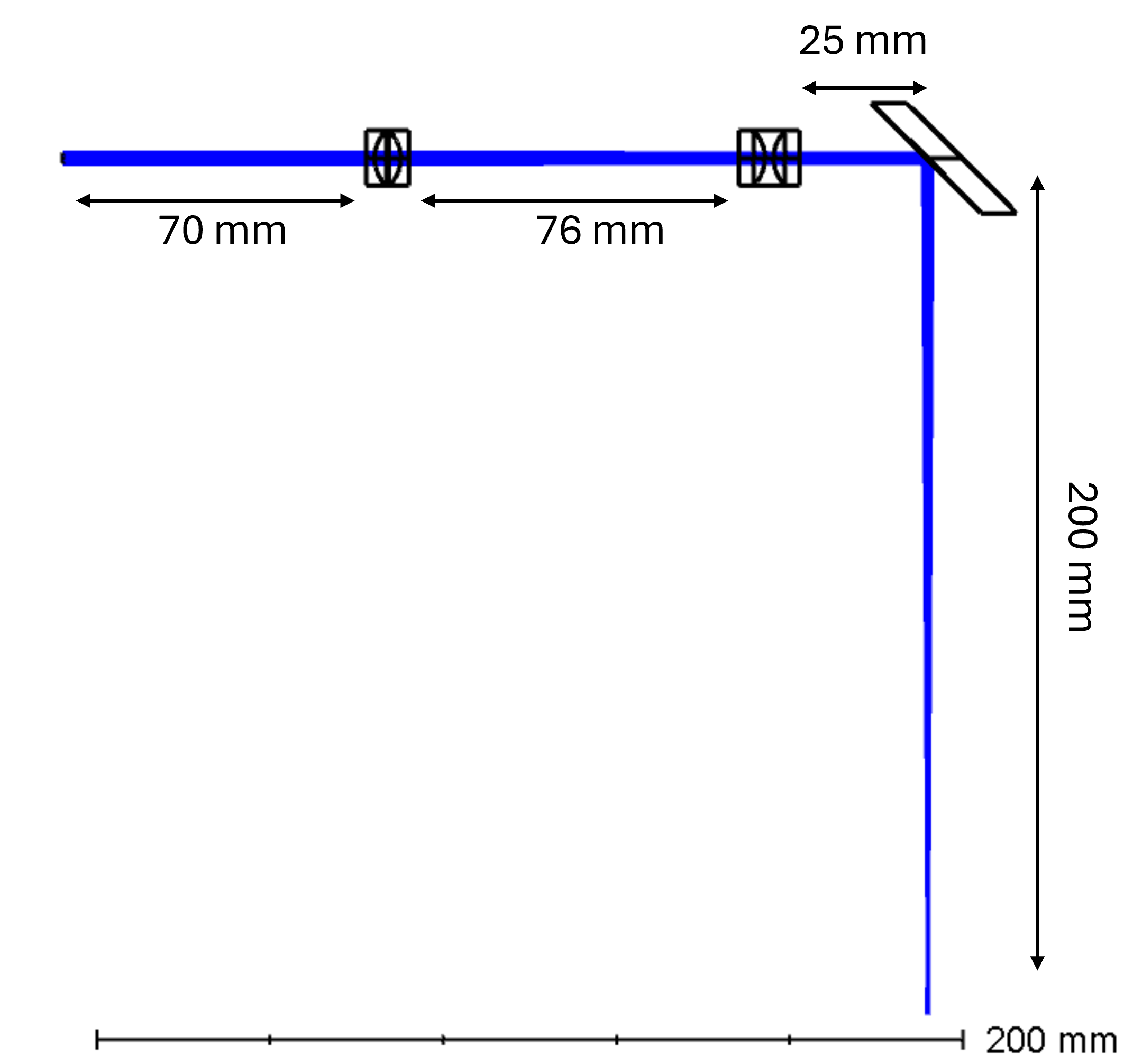}
   \end{tabular}
   \end{center}
   \caption[example] 
   { \label{fig:zemax_model} 
Zemax optical design of a pair of proposed Wyne correctors illustrating scaled layout for future testbed characterization.}
   \end{figure}

\begin{figure} [ht]
   \begin{center}
   \begin{tabular}{c} 
   \includegraphics[width=10cm]{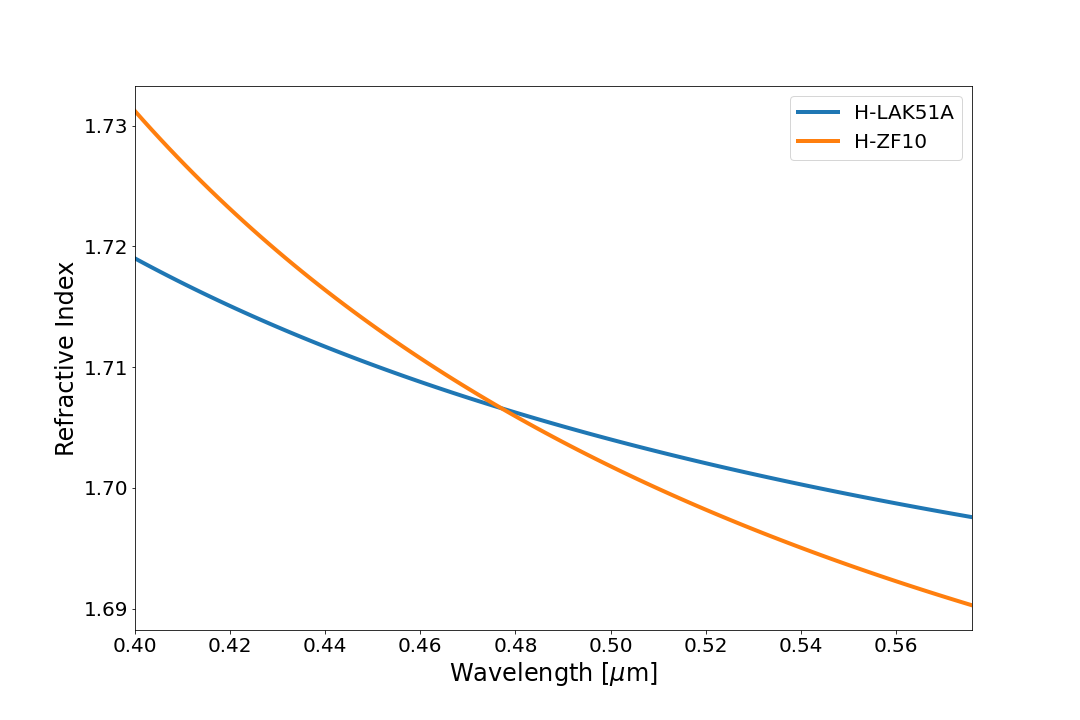}
   \end{tabular}
   \end{center}
   \caption[example] 
   { \label{fig:refractive_index} 
Refractive indices for H-LAK51A and H-ZF10 as a function of wavelength. The central wavelength $\lambda_0=480~nm$ is shown as the intersection between the two functions.}
   \end{figure}

\begin{figure}
\begin{subfigure}{.5\textwidth}
    \centering
    \includegraphics[width=.95\linewidth]{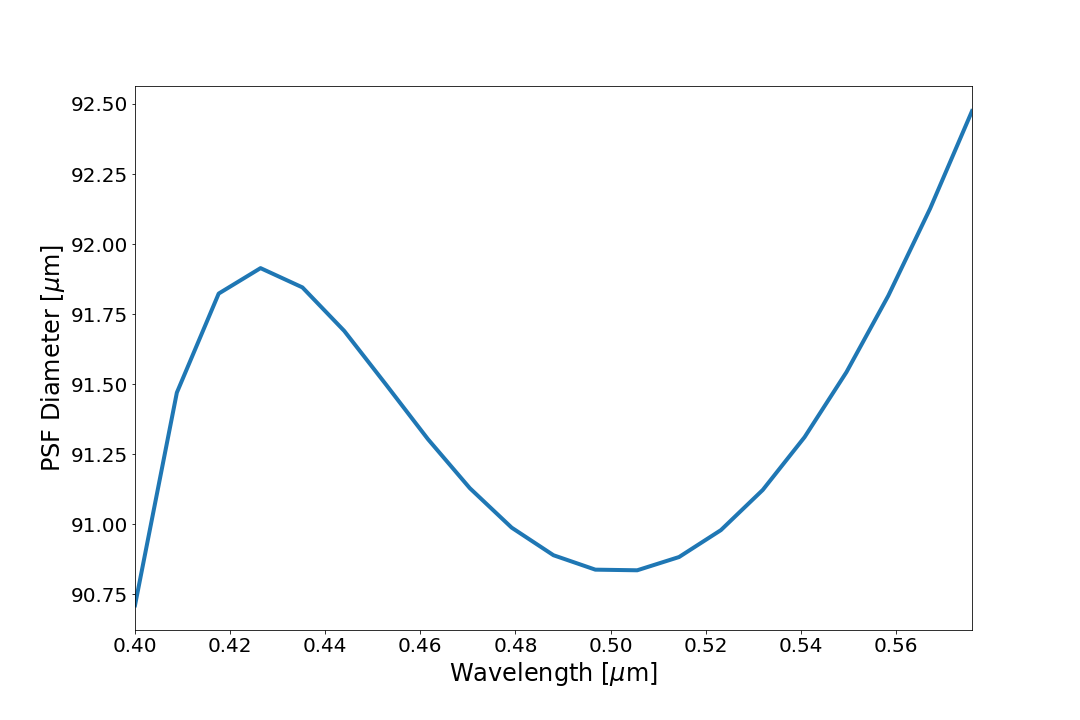}
    \caption{\label{fig:spot_dispersion}}
\end{subfigure}%
\begin{subfigure}{.5\textwidth}
    \centering
    \includegraphics[width=.95\linewidth]{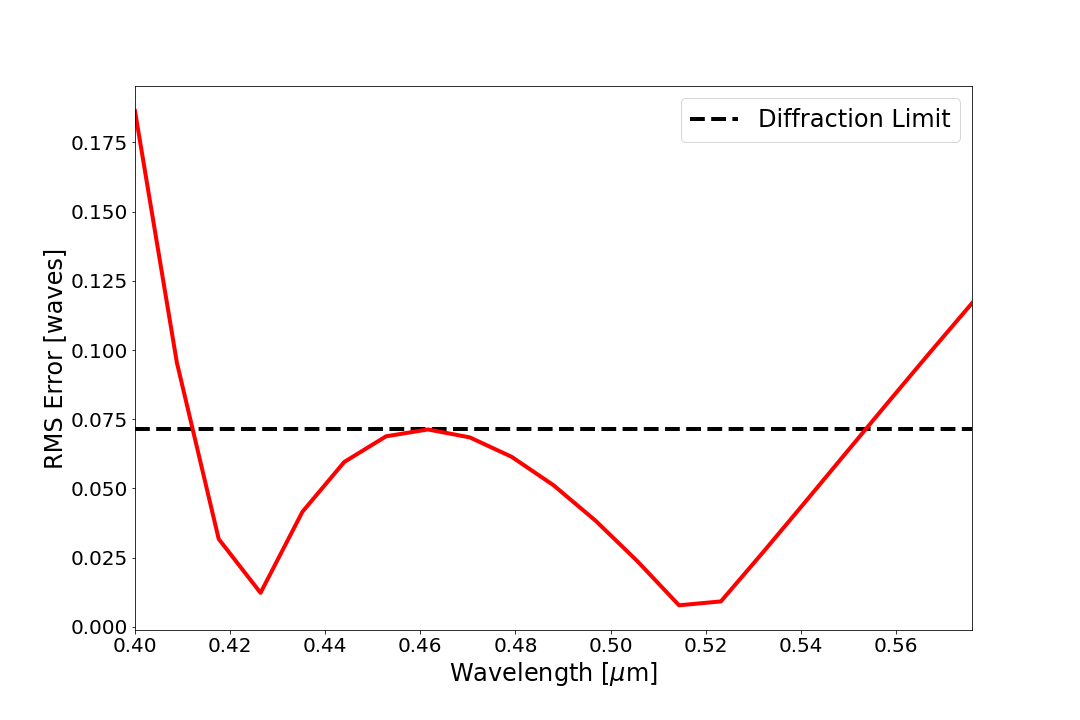}
    \caption{\label{fig:rms_wfe}}
\end{subfigure}
\caption{(a) PSF size as a function of wavelength, where the spot peak-to-valley spot dispersion is the difference between the PSF widths at the upper and lower bounds. (b) The RMS wavefront error within the designed spectral range, and the dashed line is diffraction limited RMS wavefront error.}
\end{figure}

A pair of Wyne correctors are designed, and the resulting optical layout is shown in Fig.~\ref{fig:zemax_model}. The optical elements consist of cemented triplet lenses made from H-LAK51A and H-ZF10 glass substrates, both of which exhibit the same refractive index at 480~nm, as illustrated in Fig.~\ref{fig:refractive_index}. A 2.6~mm Lyot stop generates a collimated beam, which is subsequently transformed into an $F/78$ beam using an off-axis parabolic mirror (OAP). The resulting spot dispersion and wavefront root-mean-squared (RMS) values are shown in Fig.~\ref{fig:spot_dispersion} and Fig.~\ref{fig:rms_wfe}.

The optical system is optimized by scaling the $F\#$ within the spectral range to match the PSF width at $\lambda_0=480~nm$. Subsequently, optical aberrations were minimized, making the wavefront error a critical performance metric. Without addressing these chromatic aberrations, the $F\#$ may be correct, but the resulting PSF could be too aberrated to be functional. Within a 36\% bandpass, the spot dispersion remains below 2~${\mu}$m, exceeding the initial system requirements. Table~\ref{tab:results} summarizes the system performance, indicating that the system is diffraction-limited with a 20\% bandpass and incurs a maximum wavefront error of 0.1 waves within a 36\% bandpass.

\begin{table}[ht]
\caption{} 
\label{tab:results}
\begin{center}       
\begin{tabular}{|l|l|l|}
\hline
\rule[-1ex]{0pt}{3.5ex}  \bf{Bandwidth} & 20\% & 36\%  \\
\hline
\rule[-1ex]{0pt}{3.5ex}  \bf{Spot Dispersion $[{\mu}m$]} & 1.07 & 1.76  \\
\hline
\rule[-1ex]{0pt}{3.5ex}  \bf{Wavefront Error [waves]} & $0.05\pm{0.02}$ & $0.06\pm{0.04}$  \\
\hline

\end{tabular}
\end{center}
\end{table}

\section{Conclusion}
We successfully designed Wyne correctors that increase the spectral bandwidth $>20\%$, with $<2~\mu$m of spot dispersion. Next, this optical system will be assembled and tested using the High Contrast Testbed (HCT) located at Lawrence Livermore National Laboratory to validate its performance. On HCT (F/78 beam with 6.5~$\mu$m camera pixels), $<2~\mu$m peak-to-valley dispersion allows generating a SCC dark hole (requiring < 0.3~$\lambda_{0}/D$ fringe smearing) over this 0.4 – 0.576~$\mu$m bandpass (${\Delta}\lambda/\lambda_{0}\approx36\%$) with a 5.8~$\lambda_{0}/D$ outer working angle, equivalent to a ${\Delta}\lambda_{0}/D\approx5\%$ bandpass without a Wyne corrector (i.e., a 7.2× bandwidth increase).


\acknowledgments 
This work was performed under the auspices of the U.S. Department of Energy by Lawrence Livermore National Laboratory under Contract DE-AC52-07NA27344. This document number is LLNL-PROC-865741.

\bibliography{main} 

\begin{thebibliography}{1}

\bibitem{baudoz_self-coherent_2005}
Baudoz, P., Boccaletti, A., Baudrand, J., and Rouan, D., ``The {Self}-{Coherent} {Camera}: a new tool for planet detection,'' {\em Proceedings of the International Astronomical Union}~{\bf 1},  553--558 (Oct. 2005).

\bibitem{wyne_extending_1979}
Wyne, C.~G., ``Extending the bandwidth of speckle interferometry,'' {\em Optics Communications}~{\bf 28},  21--25 (Jan. 1979).

\bibitem{delorme_focal_2016}
Delorme, J.~R., Galicher, R., Baudoz, P., Rousset, G., Mazoyer, J., and Dupuis, O., ``Focal plane wavefront sensor achromatization: {The} multireference self-coherent camera,'' {\em A\&A}~{\bf 588},  A136 (Apr. 2016).
\newblock Publisher: EDP Sciences.

\bibitem{gerard_laboratory_2022}
Gerard, B.~L., Dillon, D., Cetre, S., and Jensen-Clem, R.~M., ``Laboratory demonstration of real-time focal plane wavefront control of residual atmospheric speckles,'' {\em JATIS}~{\bf 8},  039001 (July 2022).
\newblock Publisher: SPIE.

\end{thebibliography}
\bibliographystyle{spiebib} 

\end{document}